\documentclass[preprint,journal]{vgtc}       

\ifpdf
  \pdfoutput=1\relax                   
  \usepackage{graphicx}                
  \DeclareGraphicsExtensions{.pdf,.png,.jpg,.jpeg} 
\else
  \usepackage{graphicx}                
  \DeclareGraphicsExtensions{.eps}     
\fi%

\graphicspath{{figures/}{pictures/}{images/}{./}} 

\usepackage{microtype}                 
\PassOptionsToPackage{warn}{textcomp}  
\usepackage{textcomp}                  
\usepackage{mathptmx}                  
\usepackage{times}                     
\usepackage{cite}                      
\usepackage{tabu}                      
\usepackage{booktabs}                  
\usepackage{xcolor}
\usepackage{multirow}
\usepackage{booktabs}
\usepackage{graphicx}
\usepackage{subfig}
\usepackage{enumitem}
\usepackage{float}
\usepackage[switch]{lineno}

\newcommand{\mli}[1]{\mathit{#1}}

\usepackage[ruled,linesnumbered,noresetcount]{algorithm2e}
\usepackage[utf8]{inputenc}
\usepackage{dirtytalk}
\usepackage{amsmath}
\usepackage{mathtools}
\DeclareMathOperator*{\argmin}{argmin}

\onlineid{0}

\vgtccategory{Research}
\vgtcpapertype{algorithm/technique}

\title{Sequen-C: A Multilevel Overview of Temporal Event Sequences}

\author{Jessica Magallanes, Tony Stone, Paul D Morris, Suzanne Mason, Steven Wood, and Maria-Cruz Villa-Uriol}
\authorfooter{
\item
 Jessica Magallanes and Maria-Cruz Villa-Uriol are with Department of Computer Science. E-mail: m.villa-uriol@sheffield.ac.uk.
\item
 Tony Stone and Suzanne Mason are with Centre for Urgent and Emergency Care Research. E-mail:\{ tony.stone, s.mason\}@sheffield.ac.uk.
\item Paul D Morris is with Department of Infection, Immunity and Cardiovascular Disease. E-mail: paul.morris@sheffield.ac.uk.
\item The authors above are with University of Sheffield, UK.
\item
 Steven Wood is with Sheffield Teaching Hospitals NHS Foundation Trust. E-mail: steven.wood8@nhs.net.
}

\shortauthortitle{Magallanes \MakeLowercase{\textit{et al.}}: Sequen-C: A Multilevel Overview of Temporal Event Sequences}

\abstract{Building a visual overview of temporal event sequences with an optimal level-of-detail (i.e. simplified but informative) is an ongoing challenge - expecting the user to zoom into every important aspect of the overview can lead to missing insights. We propose a technique to build a multilevel overview of event sequences, \textcolor{black}{whose granularity can be transformed across sequence clusters (\textit{vertical level-of-detail}) or longitudinally (\textit{horizontal level-of-detail})}, using hierarchical aggregation and a novel cluster data representation \textit{Align-Score-Simplify}. By default, the overview shows an optimal number of sequence clusters obtained through the average silhouette width metric – then users are able to explore alternative optimal sequence clusterings. 
The vertical level-of-detail of the overview changes along with the number of clusters, whilst the horizontal level-of-detail refers to the level of summarization applied to each cluster representation. The proposed technique has been implemented into a visualization system called Sequence Cluster Explorer (Sequen-C) that allows multilevel and detail-on-demand exploration through three coordinated views, and the inspection of data attributes at cluster, unique sequence, and individual sequence level. We present two case studies using real-world datasets in the healthcare domain: CUREd and MIMIC-III; \textcolor{black}{which demonstrate how the technique can aid users to obtain a summary of common and deviating pathways, and explore data attributes for selected patterns}. 
} 

\keywords{Temporal event sequence visualization, clustering, hierarchical aggregation, multiple sequence alignment.}

\CCScatlist{ 
 \CCScat{K.6.1}{Management of Computing and Information Systems}%
{Project and People Management}{Life Cycle};
 \CCScat{K.7.m}{The Computing Profession}{Miscellaneous}{Ethics}
}

\teaser{
  \centering
  \includegraphics[width=\linewidth]{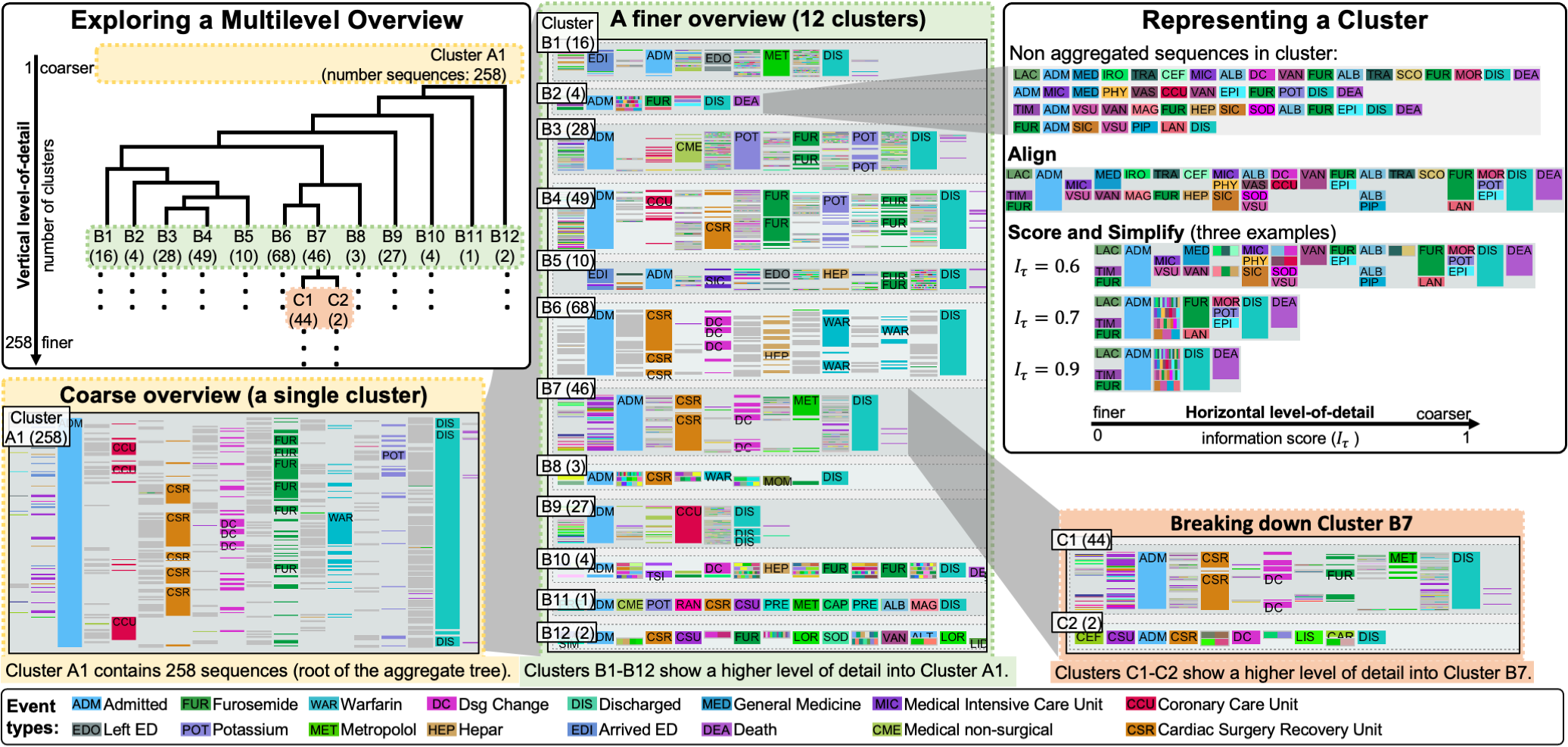}
    \caption{Screenshots of Sequen-C for a dataset with 258 event sequences illustrating the methodology. The hierarchical aggregation tree (\textit{top left}) allows changing the number of clusters shown. The vertical level-of-detail of the multilevel overview can be transformed from  coarse (\textit{bottom left}) to fine (\textit{middle} and \textit{bottom right}). Sequence clusters are represented using an Align-Score-Simplify strategy (\textit{top right}), which allows controlling the horizontal level-of-detail according to an information score.}
    \label{fig:teaser}
}

\vgtcinsertpkg

\begin{document}


\firstsection{Introduction}

\maketitle
Visual analytics of temporal event sequence data has applications in various domains such as electronic health records \textcolor{black}{(e.g.\cite{gotz2014decisionflow,guo2018visual,guo2017eventthread})} and web clickstream analysis \textcolor{black}{(e.g. \cite{liu2017patterns,wang2016unsupervised,li2020ssrdvis})}. This type of data usually presents high variability and volume. Existing visual analytic techniques obtain a visual summary (overview) of event sequences using techniques such as sequential pattern mining or sequence clustering, with the purpose of understanding common and deviating pathways. These techniques commonly follow the information-seeking mantra: \say{overview first, zoom and filter, then details on demand} \cite{shneiderman2003eyes}; which means that the starting point of the exploration is the given overview. 

Finding the optimal level-of-detail of the initial overview, simplified but informative, is an ongoing challenge. Current visualization systems limit users to a single overview as the starting point of the analysis, which can lead to missing insights unless the user drills down into specific aspects of the provided overview. Users should be allowed to seamlessly change the level-of-detail to explore alternative overviews, rather than being limited to a single one. We propose a technique to build and explore a multilevel overview of event sequences through hierarchical aggregation and the cluster data representation Align-Score-Simplify, where users can interactively transform the overview from a coarse level-of-detail to a fine one, allowing a seamless analysis of alternative overviews.

The multilevel overview presents $k$ sequence clusters retrieved from a hierarchical aggregation or \textit{aggregate tree} of event sequences. The leaves of the tree contain the original sequences and each node corresponds to a sequence cluster. \textcolor{black}{Clusters present soft patterns as opposed to hard patterns \cite{gotz2016soft}, hence order of events is not strictly enforced at this step.}
\textcolor{black}{The proposed Align-Score-Simplify data representation summarizes the sequences in a cluster (see Fig.~\ref{fig:clusterEncoding}) using Multiple Sequence Alignment (MSA) \cite{feng1987progressive} and an information score.}
\textcolor{black}{The alignment of sequences within a cluster facilitates the identification of patterns and permutations (see highlight in Fig.~\ref{fig:clusterEncoding}) in the order of events.}

The level-of-detail of the overview can be transformed \textcolor{black}{across sequence clusters and longitudinally}. The \textit{vertical level-of-detail} is proportional to the number of clusters $k$. The higher $k$ is, the finer the detail. Fig.~\ref{fig:teaser} shows the coarsest summary for $k=1$, where $k$ can go up to $258$ (the total number of sequences in the dataset). \textit{The horizontal level-of-detail} depends on the information score threshold $I_\tau$ so that the higher $I_\tau$ is, the coarser the cluster representation. 
Although users are able to interactively change the number of clusters, a set of recommended values for $k$ are suggested by analysing the Average Silhouette Width metric curve \cite{kaufman2009finding}.

We present Sequen-C, a visual analytics framework that implements the proposed technique and allows detail-on-demand exploration of sequence clusterings. To illustrate the benefits of Sequen-C, we have used two real-world datasets in the clinical domain: CUREd (Fig.~\ref{fig:caseStudyCHC}) and MIMIC-III (Fig.~\ref{fig:caseStudyMimic}). The contributions of the present work are:

\begin{itemize}[noitemsep]
  \item A technique to build and explore a multilevel overview of event sequences, from coarse to fine \textit{vertical or horizontal level-of-detail}, using hierarchical aggregation and a novel \textit{Align-Score-Simplify} cluster data representation.
  \item A novel approach to explore sequence clusterings, where a ranked set of alternative optimal number of clusters are provided.
  \item Sequen-C, a visual analytics system that allows multilevel and detail-on-demand exploration of sequences, and the inspection of attributes at cluster, unique sequence, or individual record level.
  \item \textcolor{black}{Two case studies involving domain experts in the clinical domain.}
\end{itemize}

\section{Related Work}

\subsection{Summarization of temporal event sequences}

Many techniques have been proposed to build a visual summary or overview of temporal event sequences. Early techniques display individual events in a timeline according to their time of occurrence \cite{plaisant2003lifelines}. Other ones aggregate sequences sharing the same event order using icicle plots \cite{monroe2013temporal,wongsuphasawat2011lifeflow}, Sankey diagrams \cite{wongsuphasawat2012exploring}, or transition matrices \cite{zhao2015matrixwave}. They visualize all event permutations, complicating the interpretability as volume and variability increase. Scalability has been addressed by obtaining an overview of relevant (e.g. frequent) sub-sequences through sequential pattern mining (SPM) \cite{liu2017patterns,kwon2016peekquence,vrotsou2009activitree}. 
Despite SPM being able to summarize datasets with high volume and variability, hierarchical exploration in the level-of-detail is not possible.

Frequence \cite{perer2014frequence} builds a coarse level-of-detail overview of frequent patterns, from which sub-patterns with finer level-of-detail are interactively mined, however, this method relies on existing event type hierarchical categories. Coreflow \cite{liu2017coreflow} obtains an overview of branching patterns using a three-step recursive algorithm, which allows its hierarchical exploration by expanding branches into sub-branches; nevertheless, the interaction is constrained to the clicked branches and no support is provided hinting users towards alternative levels-of-detail which might provide valuable insights. Other approaches merge and replace events \cite{monroe2013temporal}, define temporal queries \cite{fails2006visual,monroe2013challenges}, define milestone events \cite{gotz2014decisionflow} or use regular expressions \cite{zgraggen2015s,cappers2018exploring}. 

User-driven operations applied to the overview are key for visual exploratory data analysis. However, the choice of the initial overview remains crucial \cite{elmqvist2009hierarchical}. Our paper proposes a methodology to build a multilevel overview of temporal event sequences, which can be transformed from coarse to fine level of details, using sequence clustering and multiple sequence alignment.

\subsection{Event and Sequence Clustering}
\textbf{Event clustering}: Outflow \cite{wongsuphasawat2012exploring} visualizes common pathways using a Sankey-like diagram, and to reduce visual clutter, events are clustered according to their outcome. However, the clustering is limited to events within the same layer. Gotz \textit{et al.}~\cite{gotz2019visual} propose a technique for dynamic hierarchical aggregation of events, where users can choose alternative groupings within a hierarchy of events, according to their correlation with outcome. Scribe Radar \cite{wongsuphasawat2014using} builds a hierarchy of event types based on their frequency and a six-level naming hierarchy. These approaches rely on an existing hierarchy of event types (e.g. hierarchical classification of clinical codes), which do not necessarily exist in every domain or dataset. Our paper focuses in the hierarchical aggregation of sequences rather than event types, and we build the hierarchy without using \textit{a priori} knowledge.

\textbf{Sequence clustering}: Wang \textit{et al.}~\cite{wang2016unsupervised} build an overview of user behavioral patterns by clustering users with a similar clickstream history, and are visualized using a Packed Circle view. The overview successfully captures the hierarchy of nested clusters, however, the event sequences are not included. Wei \textit{et al.}~\cite{wei2012visual} build an overview of clickstream clusters mapped onto a 2D plane. The sequential information is visually encoded but the separation of different clusters is difficult to interpret. Another strategy is to explicitly encode the event sequences stacked and grouped by cluster, comparing them side to side \cite{cadez2003model,stragier2019data}. \textcolor{black}{Li et al.\cite{li2020ssrdvis} uses a density-based clustering method and visualizes clusters using a Voronoi map and a pattern graph. The latter two do not facilitate} the comparison of sequences within or amongst clusters. Treemaps are commonly used to visualize hierarchical clusters \cite{makanju2008logview,gotz2011visual}, but this type of view does not encode sequential information. \textcolor{black}{Other approaches to cluster sequences include tensor analysis \cite{guo2017eventthread} and Hidden Markov Models \cite{kwon2020dpvis} with a focus on obtaining evolution patterns.}

Vasabi and Sequence Synopsis are the techniques most similar to this work. Vasabi \cite{nguyen2019vasabi} builds an overview of sequence clusters by first extracting the most common events (e.g. tasks) in the dataset and then clustering sequences using those events as features of the clustering. The technique successfully extracts and represents a fixed number of sequence clusters. However, their cluster representation does not allow to identify event permutations or see the events that were omitted in the event extraction step, and the number of clusters cannot be changed. Sequence Synopsis \cite{chen2018sequence} clusters event sequences based on the minimum description length, where clusters are represented by a sequential pattern and a set of corrections; as indicated by Chen \textit{et al.}~\cite{chen2018sequence}, a potential drawback of their cluster representation is that missing events are not explicitly encoded and that scalability could be improved by supporting hierarchical visual summary (e.g. explore alternative number of clusters). To the best of our knowledge, there is no existing technique to explore different sequence clusterings that at the same time provides an interpretable representation of the sequences in a cluster - the present work aims to tackle this problem.

\subsection{Sequence alignment}
Alignment is a common analytic strategy in temporal event data that allows to explore the events happening right before and after a given event \cite{du2016coping}. Existing techniques usually allow alignment by a single event \cite{monroe2013temporal,chen2018sequence, wongsuphasawat2012exploring}, \textcolor{black}{and more recently by two \cite{zhang2019evaluating, zhang2018idmvis}} or multiple \cite{cappers2018exploring} events. Multiple Sequence Alignment (MSA) \cite{feng1987progressive} was initially proposed to align biological sequences and understand how they relate to each other. Bose \textit{et al.}~\cite{bose2010trace} applies this algorithm to temporal event sequences to find common behaviour and deviations in a process, with applications in domains such as clinical workflows \cite{zhou2017evaluation,bouarfa2012workflow}. MSA has also been used to obtain a consensus sequence (i.e. a set of common events) to represent a set of event sequences \cite{goodstadt2001chroma, wang2003segid, lee2003generating, di2020sequence}. We apply an MSA approach to represent the common events in a sequence cluster to allow comparison of commonalities and deviations within and across clusters.

\section{A multilevel overview of event sequences}\label{section:methodology}

We propose a technique to build and explore a multilevel overview of event sequences through hierarchical aggregation. A multilevel overview is a visual summary which can be interactively transformed from coarse to fine level-of-detail \cite{elmqvist2009hierarchical}. Our overview displays a  number of sequence clusters retrieved from an \textit{aggregate tree}; where each sequence cluster is represented with the steps Align-Score-Simplify. 

The overview can be interactively transformed vertically and horizontally. The vertical level-of-detail is controlled with the number of clusters retrieved from the tree.  Fig.~\ref{fig:aggregateTree}-C shows how the higher in the hierarchy (i.e. smaller number of clusters), the coarser the overview; whereas the lower in the hierarchy (i.e. larger number of clusters), the finer the details. The horizontal level-of-detail refers to the level of simplification of each cluster representation according to its information score (Fig. ~\ref{fig:clusterEncoding}). The initial overview shows the best number of clusters according to the Average Silhouette Width metric\cite{rousseeuw1987silhouettes}, but we also offer a set of alternative values that might provide valuable  overviews.

\subsection{Building the aggregate tree}\label{sub:buildTree}
\begin{figure*}[t]
   \includegraphics[height=0.45\textheight]{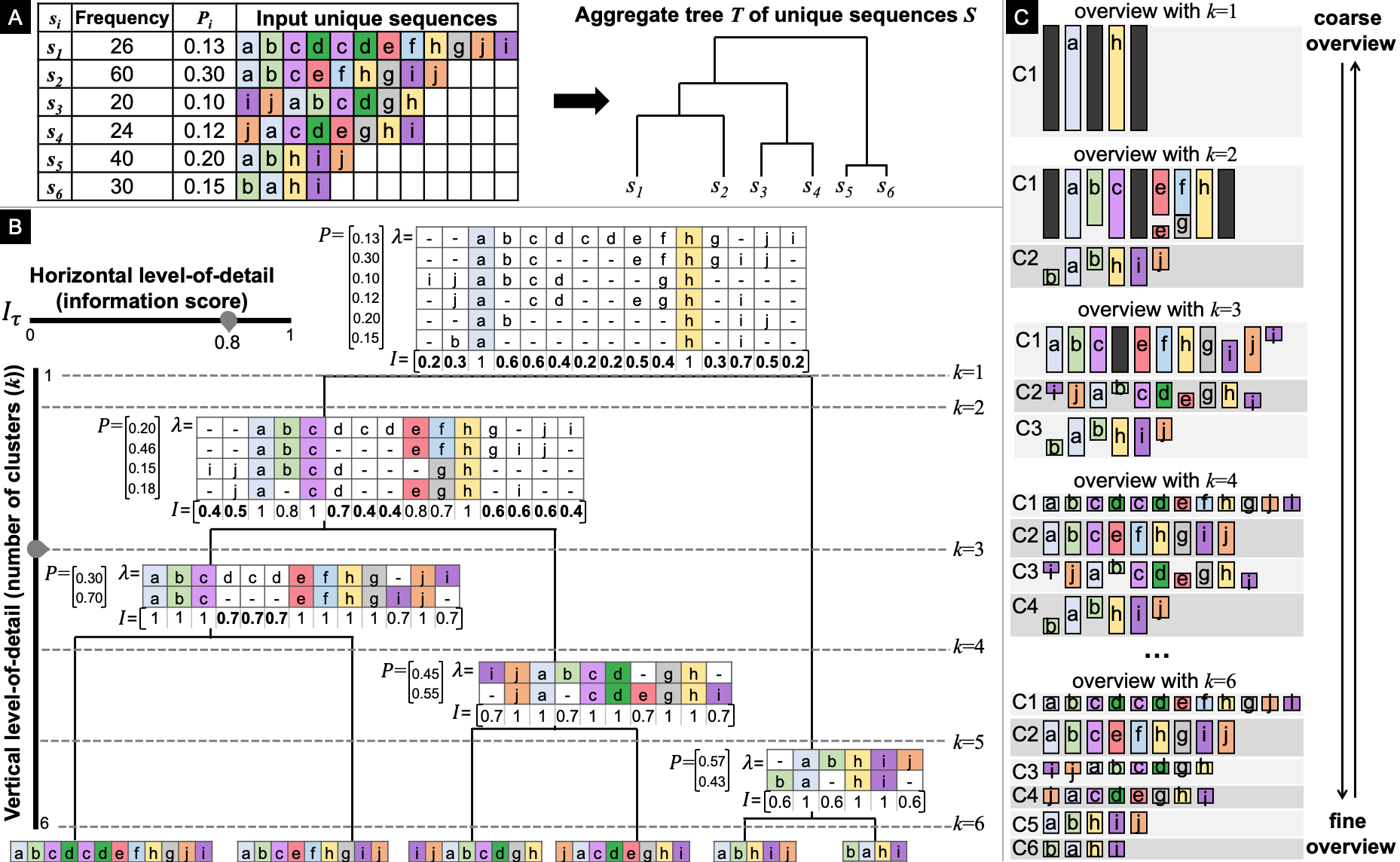}
   \caption{(A) Building  aggregate tree $T$ for input unique sequences $S=\{s_1, ...s_6)\}$. (B) Each node in $T$ has an alignment matrix $\lambda$ for its child sequences, a row-wise probabilities vector $P$, and a column-wise information score vector $I$. Two or more consecutive columns in $\lambda$ with $I_j < 0.8$ are not coloured. (C) Multilevel overviews for a range of number of clusters $k$ retrieved from $T$, where black blocks represent merged columns.}
   \label{fig:aggregateTree}
\end{figure*}

To build the aggregate tree from the input unique temporal event sequences, we use a bottom-up aggregation approach \cite{aggarwal_reddy_2014}. Every input unique sequence starts in a single cluster, then pairs of similar clusters are iteratively aggregated until a single cluster is obtained. The leaves of the final tree contain the input unique sequences, each internal node in the tree contains an aggregated representation of their child nodes, and the root node aggregates the whole dataset (see Fig.~\ref{fig:aggregateTree}-B). 

Alg.~\ref{alg:BuildAggregateTree} shows how to build the aggregate tree $T$ from the list of unique sequences $S=\{s_1,s_2,...,s_N\}$, being $N$ the number of input unique sequences in the dataset. In the last iteration, $T$ is the root node of a binary tree, and all possible sub-trees can be obtained by recursively retrieving its left and right children until reaching the leaves.

First, the distance matrix $d$ is initialized. This  $N \times N$ matrix provides the pairwise distances of all the input sequences in $S$.   $\mli{D_{qgram}}$ computes the pairwise distance of two sequences as the cosine distance of their q-gram profiles \cite{ukkonen1992approximate}, where the q-gram profile of a sequence is the vector of all sub-sequences of $q$ consecutive events. The implementation of the \texttt{stringdist} package in R \cite{van2014stringdist} was used, with $q=1$. This allows clustering sequences according to the count of shared events regardless of their permutations in order. For example, the distance between sequences ``abcde" and ``deabc" is zero and both sequences are likely to end up in the same cluster. \textcolor{black}{In case a stricter ordering of events had to be considered, $D_{qgram}$ should be replaced by another distance metric (e.g. Levenshtein edit distance \cite{levenshtein1966binary}).}

A node in $T$ is defined as $\mli{Node}(l,r,\lambda,\alpha)$; where $l$ and $r$ are the left and right children, $\lambda$ the alignment of the sequences at that node, and $\alpha$, the simplified $\lambda$ used to represent the cluster. 
The set $T$ is initialised with as many leaves as input sequences (line 3). For each $s$ in $S$, a leaf is  $\mli{Node}(l=\emptyset,r=\emptyset,\lambda=\alpha=s)$; with no children, and where the alignment $\lambda$ and cluster representation $\alpha$ are the sequence $s$ itself.

The iterative process stops when $T$ contains a single node (line 4). For each iteration a new node $n_{a \cup b}$ is created by merging the closest pair of nodes $(n_a,n_b)$, and $d$ and $T$ are updated according to $n_{a \cup b}$. 
The closest pair of nodes $(n_a,n_b)$ is that for which the value in $d$ is the minimum (line 5). Then $\texttt{aggregate}$ returns a new node $n_{a \cup b}$ whose left child is $n_a$ and right child is $n_b$, with alignment $\lambda$ and a summarized representation $\alpha$ of the sequences in $n_a$ and $n_b$. Alg.~\ref{alg:VisualRepCluster} outlines how to build the cluster data representation of the newly aggregated node $n_{a \cup b}$. 
$d$ and $T$ are updated by adding the new node $n_{a \cup b}$ and removing the nodes $n_a$ and $n_b$ used for the aggregation (lines 7-10). The distance matrix $d$ is updated by inserting the pairwise distance from the new node $n_{a \cup b}$ to all nodes in $T$ (line 7), and by removing from $d$ the $i_{th}$ row and $j_{th}$ column containing $n_a$ and $n_b$. The method $D_\mli{cqgram}$ (line 7) computes the distance between two clusters using the average agglomeration method \cite{aggarwal_reddy_2014} defined as the average of all pairwise distances between the sequences in both clusters.

\begin{algorithm}[h]
    \KwData{input unique sequences $S=\{s_1,s_2,...,s_N\}$} 
    \KwResult{aggregate tree $T$}
    \SetKwFunction{FMain}{buildAggregateTree}
    \SetKwProg{Fn}{Function}{:}{\KwRet $T$}
    \Fn{\FMain{$S$}}{
        \tcc{initialize pairwise distance matrix $d$}
        $d[i,j] = D_\mli{qgram}(s_i,s_j);\;\;\;\forall$ pairs, $s_i,s_j \in S$\\
        \tcc{initialize $T$ as a list of leaf nodes}
        $T=\{\mli{Node}(l,r,\lambda,\alpha) \mid l=\emptyset,r=\emptyset,\lambda=\alpha=s, \;\;\forall\; s \in S\}$\;
        \tcc{loop until one node remains (root node)}
        \While{$|T| > 1$}{
        
            \tcc{aggregate the closest pair of nodes}
            
            $(n_a,n_b) = \argmin d[n_i,n_j];\;\;\;\forall$ pairs, $n_i,n_j \in T $

            $n_{a \cup b} = \textbf{\texttt{aggregate}}(n_a,n_b)$\;
            
            \tcc{update distance matrix}
            
            $d[n_{a \cup b},n] = d [n,n_{a \cup b}] = D_\mli{cqgram}(n_{a \cup b},n); \;\;\forall\; n \in T$\\
            
            remove distances containing $n_a$ and $n_b$ from $d$\;
            
            \tcc{update nodes set $T$}
             
            remove $n_a$ and $n_b$ from $T$\;
            add $n_{a \cup b}$ to $T$\;
        }
    }
    \caption{Build aggregate tree}
    \label{alg:BuildAggregateTree}
\end{algorithm}
\begin{algorithm}[h!]
    \KwData{nodes $n_a$ and $n_b$}
    \KwResult{aggregated node $n_{a \cup b}$}
    \SetKwFunction{FMain}{aggregate}
    \SetKwProg{Fn}{Function}{:}{\KwRet $n_{a \cup b}$}
    \Fn{\FMain{$n_a,n_b$}}{
        \tcc{\textbf{Align}: compute $\lambda$ from children nodes}
        $\lambda = \mli{\texttt{MSA}}(n_a.\lambda, n_b.\lambda)$\;
        \tcc{\textbf{Score}: column-wise information score $I$}
        \For{$j \gets 1$ \textbf{to} $m$}{
            Compute $I_j$ according to Eq.~\ref{eq:infoScore} and Eq.~\ref{eq:entropy}
        }
        \tcc{\textbf{Simplify}: collapse columns based on $I$}
        $\mli{listRemove}=\emptyset$\;
        \For{$j \gets 1$ \textbf{to} $m-1$}{
            \If{$I_j$ $<$ $I_\tau$ and $I_{j+1}$ $<$ $I_\tau$}{
                \For{$i \gets 1$ \textbf{to} $n$}{
                    $\lambda_{i,j+1}=\texttt{concatenate}(\lambda_{i,j},\lambda_{i,j+1})$\;
                }
                add $j$ to $\mli{listRemove}$\;
            }
        }
        \tcc{assign the simplified alignment to $\alpha$}
        $\alpha=$ delete columns in $listRemove$ from $\lambda$\;
        \tcc{create new node $n_{a \cup b}$}
        $n_{a \cup b}=\mli{Node}({n_a},{n_b},\lambda,\mli{\alpha})$\;
    }
    \caption{Data representation of a cluster}
    \label{alg:VisualRepCluster}
\end{algorithm}

\subsection{Cluster data representation: Align-Score-Simplify}\label{sub:dataRepresentation}
\begin{figure*}[t]
   \includegraphics[width=\textwidth]{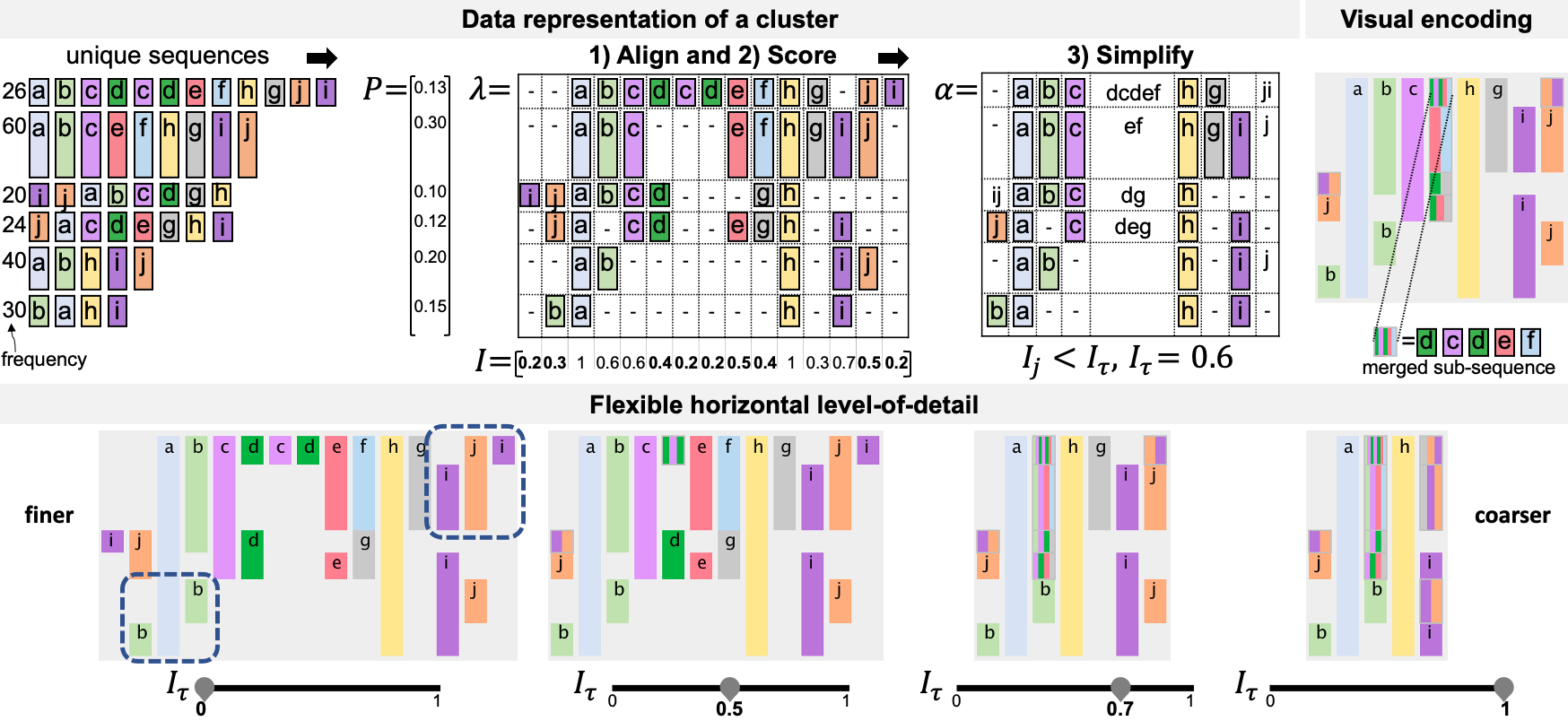}
   \caption{Cluster data representation (\textit{top left}) and visual encoding (\textit{top right}) for $S=\{s_1,...s_6\}$ and $I_\tau=0.6$ using the Align-Score-Simplify strategy. The Simplify step produces the simplified alignment matrix $\alpha$. Using $\alpha$, the visual encoding is built and event types in the merged sub-sequences are represented using ordered colored bars. Changing $I_\tau$ (\textit{bottom}) transforms the horizontal level-of-detail. \textcolor{black}{Note how this shows the most common event orderings (e.g. a-b-c-h) and permutations in the order (e.g. b-a and j-i highlighted with dashed lines).}}
   \label{fig:clusterEncoding}
\end{figure*}

To represent a cluster, we propose the steps Align-Score-Simplify (Fig.~\ref{fig:clusterEncoding}).
The sequences in a cluster are  aligned using Multiple Sequence Alignment (MSA) \cite{feng1987progressive}. Then an information score is computed for each column in  $\lambda$. And last, the columns in $\lambda$ with an information score below a threshold $I_\tau$ are merged and simplified.

\subsubsection{Align}\label{sub:align}

To perform the Align and obtain the alignment matrix $\lambda$, the input unique sequences in $S$ are formatted as sequences of characters, where each character represents an event type. All elements in $\lambda$ are either single characters or gaps (--). The MSA algorithm inserts gaps (--) in those input sequences to maximize the number of equal events column-wise. To achieve this, two costs are used: gap open penalty (when inserting gaps), and gap substitution score (encouraging equal column-wise events).

We use the progressive approach \cite{feng1987progressive} to carry out the multiple alignment of sequences, which iteratively constructs a series of pairwise alignments by following a tree that represents the similarity between sequences, where the alignment of a node is built using the alignment of its child nodes (Fig.~\ref{fig:aggregateTree}-B). Alignments can be constructed over a pair of sequences, a sequence and an alignment, or a pair of alignments. For a given set of sequences $S$, the $N \times M$ alignment matrix $\lambda$ contains all the sequences in $S$ aligned, being $N$ the number of input unique sequences and $M$ the length of the final alignment  \cite{bose2010trace}. The method MSA (line 14) returns  $\lambda$ for the sequences in the new node, computed using the alignments of its children $n_a.\lambda$ and $n_b.\lambda$. In this paper, the algorithm was written in R and  based on the Python library \texttt{scikit-bio} \cite{scikit-bio}, with $\mli{gap\_open\_penalty}=0.8$, $\mli{match\_score}=3$ for equal events, and $\mli{mismatch\_score}=-1$ for non equal events.

\newcommand{\STAB}[1]{\begin{tabular}{@{}c@{}}#1\end{tabular}}
\renewcommand{\arraystretch}{1.05}
\begin{table*}[]
\begin{tabular}{|c|c|c|c|c|c|c|c|c|c|c|}
\hline
\multirow{4}{*}{\STAB{\rotatebox[origin=c]{90}{\textbf{Dataset}}}} & \multirow{4}{*}{\textbf{Percentage}} &   \multicolumn{2}{c|}{\textbf{No.~sequences}} &  \multirow{4}{*}{\textbf{\begin{tabular}[c]{@{}c@{}}No. event\\ types\end{tabular}} } & \multicolumn{2}{c|}{\textbf{Length of sequences}} & \multicolumn{4}{c|}{\textbf{Execution time (s)}}  \\
        \cline{3-4}\cline{6-11} 
& & \multirow{3}{*}{\textbf{Individual}} & \multirow{3}{*}{\textbf{Unique}} &  & \multirow{3}{*}{\textbf{Average}} & \multirow{3}{*}{\textbf{Maximum}} &  \texttt{build} & \multicolumn{2}{c|}{\texttt{aggregate}} & \multirow{3}{*}{\textbf{Total}}\\ 
\cline{9-10} 
& &  &  &  &  &  & \texttt{AggregateTree}  & \multirow{2}{*}{\textbf{Align}} &  \multirow{2}{*}{\textbf{Score \&}} & \\ 
\rule{0pt}{9pt}
 & &  &  &  &  &  &  &  &  \textbf{Simplify} & \\ 
 \hline
\multirow{4}{*}{\STAB{\rotatebox[origin=c]{90}{CUREd}}} & 25\% & 20,124 & 240 & 11 & 9.96 & 13 & 0.0 & 9.3 & 10.1 & 19.4 \\ \cline{2-11}
& 50\% & 20,901 & 481 & 11 & 13.07 & 19 & 0.1 & 59.1 & 30.9 & 90.1 \\ \cline{2-11}
& 75\% & 21,520 & 722 & 11 & 16.24 & 28 & 0.2 & 225.5 & 74.5 & 300.2 \\ \cline{2-11}
& \textbf{100\%} & \textbf{21,805} & \textbf{962} & \textbf{11} & \textbf{22.70} & \textbf{177} & \textbf{0.4} & \textbf{3984.7} & \textbf{301.0} & \textbf{4286.0} \\ \hline
\multirow{4}{*}{\STAB{\rotatebox[origin=c]{90}{MIMIC-III}}} & 25\% & 442 & 328 & 158 & 6.07 & 8 & 0.1 & 3.5 & 9.7 & 13.2 \\ \cline{2-11}
& 50\% & 770 & 656 & 242 & 7.31 & 10 & 0.2 & 16.9 & 35.6 & 52.7 \\ \cline{2-11}
& 75\% & 1,097 & 983 & 338 & 8.45 & 12 & 0.5 & 73.4 & 86.9 & 160.8 \\ \cline{2-11}
& \textbf{100\%} & \textbf{1,425} &\textbf{ 1,311} & \textbf{448} & \textbf{10.67} & \textbf{96} & \textbf{1.5} & \textbf{995.4} & \textbf{317.5 }& \textbf{1314.3} \\ \hline
\end{tabular}
\caption{\textcolor{black}{Characteristics of datasets used in the case studies (in bold), and time performance for algorithms $\texttt{buildAggregateTree}$ and $\texttt{aggregate}$ for different subsets of the data. The time for $\texttt{buildAggregateTree}$ does not include the function $\texttt{aggregate}$, which is broken down according to the Align, Score and Simplify steps. Experiments were run on a MacBook Pro, 2.9 GHz Quad-Core Intel Core i7, 16GB (Appendices A and B).}}
   \label{table:performanceTable}
\end{table*}

\subsubsection{Score}

Once the alignment matrix $\lambda$ for all the input unique sequences in a cluster (or node) is obtained, $\lambda$ is simplified into the simplified alignment matrix $\alpha$. 
To do this, an \textit{information score} $I_j$ for each column $j$ in $\lambda$ is computed based on \cite{bose2010trace}:

\begin{equation}
I_j = 1-\dfrac{E_j}{\log_2(|A| + 1)}
\label{eq:infoScore}
\end{equation}
being $A$ the set of unique event types in the alignment matrix $\lambda$, $|A|$ the length of $A$, and $E_j$ the entropy of the event types in that column: 
\begin{equation}
E_j = \sum_{a \in A_j \cup \{-\}} 
\begin{cases*}
  -P_a \log_2\left(\dfrac{P_a}{G_j}\right), & \textrm{if} a = `--',\\
  -P_a \log_2(P_a), & otherwise.
\end{cases*}
\label{eq:entropy}
\end{equation}

$\mli{G_j}$ is the count of gaps in column $j$, $P_a$ is the probability of the event type $a$ in that column, and $A_j$ is the set of unique event types in column $j$. To avoid $I_j$ from becoming negative, when $E_j > \log_2(|A| + 1)$, $E_j=\log_2(|A| + 1)$. The probability $P_a$ is computed as the sum of probabilities of the unique sequences to which $a$ belongs to, defining the probability of a unique sequence as its frequency divided by the total frequency of the cluster. Fig.~\ref{fig:aggregateTree} illustrates how $P$ and $I$ are calculated for each alignment matrix $\lambda$ in the aggregate tree $T$.

The information score provides a measure of how homogeneous a column is, with values in the range $0 \leq I_j \leq 1$. If a column contains mostly a single type of event, $I_j$ is closer to one, while if it mostly contains gaps or many distinct event types, $I_j$ is closer to zero. The vector $I$ is computed according to Eqs.~\ref{eq:infoScore} and \ref{eq:entropy} (lines 15-17, Alg.~\ref{alg:VisualRepCluster}).

\subsubsection{Simplify}\label{sub:simplify}

To simplify an alignment matrix $\lambda$, columns with a relatively low information score are iteratively merged to obtain the simplified cluster representation matrix $\alpha$ (lines 19 to 26, Alg.~\ref{alg:VisualRepCluster}). The matrix $\alpha$ is $N \times M'$, where $M'$ is the length of the simplified alignment, being $M' \leq M$. 

Alg.~\ref{alg:VisualRepCluster} shows how given a pair of consecutive columns with information score $I_j$ and $I_{j+1}$ below a threshold $I_\tau$ (line 20), the characters in $\lambda_{i,j}$ are concatenated to the beginning of $\lambda_{i,j+1}$. Such concatenation is repeated for each row in the alignment for the selected columns, then column $j$ is added to the list of columns to be removed ($\mli{listRemove}$). The matrix $\alpha$ is the resulting simplified alignment and contains the same columns as $\lambda$ except for the columns in $\mli{listRemove}$, i.e. the columns categorized as candidates for a horizontal merge (line 27). Note that $\alpha$ will have elements that are a concatenation of characters, whereas $\lambda$ only contains single characters. Those elements in $\alpha$ that are a concatenation of characters represent the merged sub-sequences.

Fig.~\ref{fig:clusterEncoding} shows an example about how Simplify  works. The events in $\lambda$ (represented as characters) in columns 1 to 2 and 6 to 10 are row-wise merged into a single position in the final simplified matrix of alignment $\alpha$. The visual encoding of these row-wise merged events is further explained in Section \ref{section:system}. Alg.~\ref{alg:VisualRepCluster} finishes by creating the new node $n_{a \cup b}$, assigning $n_a$ and $n_b$ as its child nodes, $\lambda$ as its alignment, and $\alpha$ as its data representation (line 28). Note that the original alignment matrix $\lambda$ is also kept as it is used to build the alignment of subsequent nodes.

\subsubsection{Multilevel data representation}

The proposed data representation allows to explore the overview vertically and horizontally. The \textbf{vertical level-of-detail} is proportional to the number of clusters $k$ in the overview. The larger $k$ is, the finer the overview, where $1 \leq k \leq N$, being $N$  the number of input unique sequences. For example, Fig.~\ref{fig:aggregateTree}-B shows how as the aggregate tree is cut at a higher level in the hierarchy (e.g. $k=1$ or $k=2$), clusters have a higher intra-cluster variation so the number of merged columns increase, resulting in a coarser overview. As the tree is cut at a lower level in the hierarchy (e.g. $k=4$ or $k=5$), clusters have a lower intra-cluster variation so the column-wise information score gets closer to 1, resulting in a finer overview. Ultimately, when one cluster contains a single sequence ($k=6$), all columns in the alignment matrices have an information score of 1, showing an overview with the highest level-of-detail possible. 

The \textbf{horizontal level-of-detail} of the overview depends on the threshold $I_\tau$, where $0 \leq I_\tau \leq 1$. The larger $I_\tau$ is, the coarser the representation of clusters becomes horizontally. Fig.~\ref{fig:clusterEncoding} shows the representation for an example cluster for several $I_\tau$ values, when $I_\tau=0$ no events are merged showing full detail and as it moves towards $I_\tau=1$ the number of merged events increase.

\subsection{Finding optimal overviews}\label{sub:optimalOverviews}

Users are able to explore all clustering combinations in the aggregate tree. The average silhouette width metric \cite{kaufman2009finding} measures the quality of a clustering to find the optimal number of clusters which reflect homogeneous and well-separated distinct groups. We use this metric to suggest a set of optimal number of clusters, which result in a set of overviews with an optimal vertical level-of-detail. For a given number of clusters $k$, the \textit{average silhouette width} $\bar{z}(k)$ is defined as the mean $z(s)$ of all the elements in the dataset, where $z(s)$ is the \textit{silhouette value} of the element $s$. In our case, $s$ is each of the input unique sequences used to build the aggregate tree. As defined by Rousseeuw \cite{rousseeuw1987silhouettes}, $z(s)$ is given by: $z(s) = \frac{v(s) - u(s) }{\max(u(s),v(s))}$, where $u(s)$ is the average distance between element $s$ and the other elements in the same cluster, $v(s)$ is the average distance between $s$ and the elements in the nearest cluster (neighboring cluster), and $-1\leq z(s) \leq 1$. Kaufman and Rousseeuw \cite{kaufman2009finding} suggest that the most optimal $k$ is that one for which $\bar{z}(k)$ is the largest (global maxima). In some cases, the most optimal $k$ might still be too many or too few clusters for the user. To provide a balance between number of clusters and quality of clustering, we propose to obtain the peaks (local maxima) in the $\bar{z}(k)$ function. A set of optimal overviews are indicated by the global and local maxima in $\bar{z}(k)$. These will indicate relative good partitioning of the sequences and therefore provide a good visual overview.

\begin{figure*}
   \includegraphics[width=\textwidth]{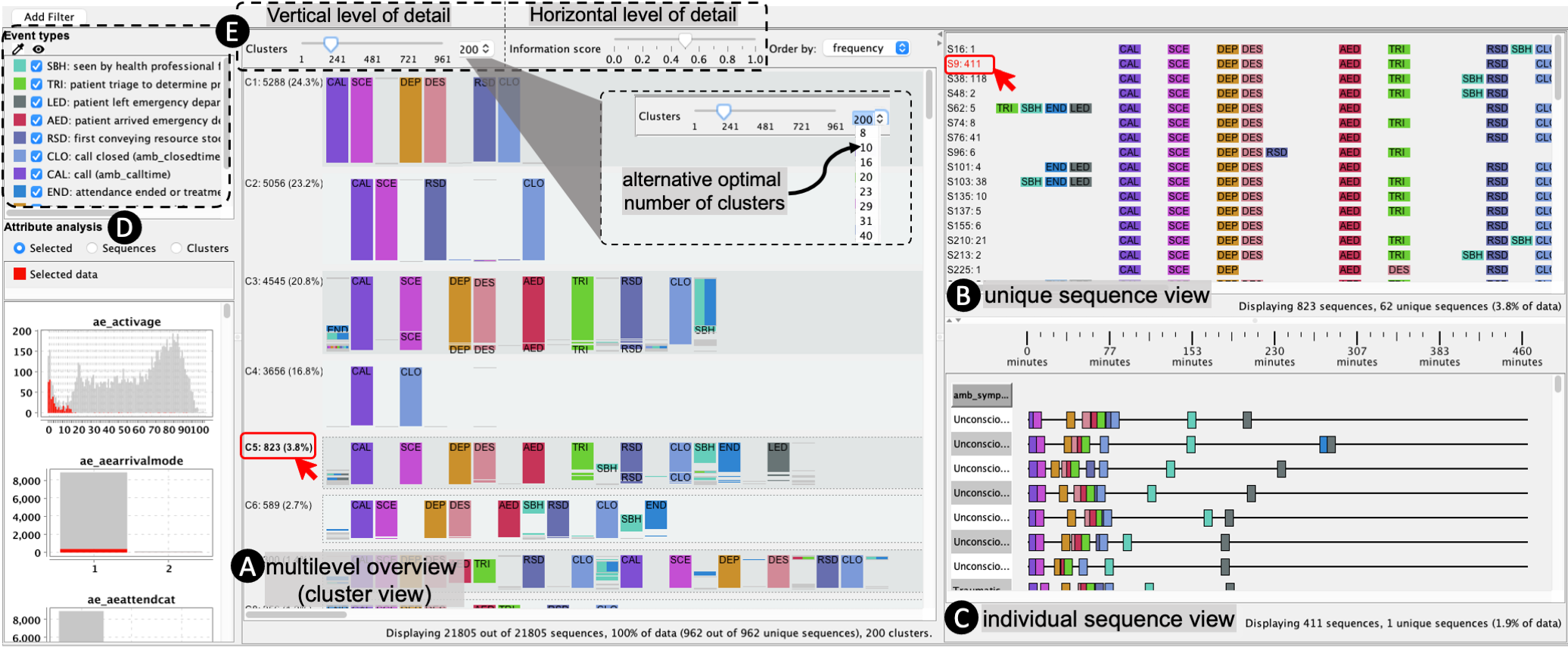}
   \caption{Sequen-C visualization system (CUREd dataset).
   In the multilevel overview (A), cluster C5 is selected and its unique sequences are shown in the unique sequences view (B), where unique sequence S9 is selected, showing its 411 individual sequences in the individual sequences view (C). The attribute analysis (D) shows how attributes of the selected data relate to the whole dataset. (E) highlights some of the available controls. }
   \label{fig:coordinatedViews}
\end{figure*}

\subsection{\textcolor{black}{Complexity analysis}}
\textcolor{black}{The time complexity of $\texttt{buildAggregateTree}$ (Alg.~\ref{alg:BuildAggregateTree}) is $O(Nnl^2)$, where $N$ is the number of input unique sequences, $l$ the maximum sequence length, and $n$ the average number of sequences per node. The function $\texttt{aggregate}$ is repeated $N-1$ times, being the Align step ($MSA$) the most time consuming ($O(nl^2)$). Table~\ref{table:performanceTable} compares the time for several subsets of the data used in our case studies, and as observed, the maximum sequence length in the dataset highly impacts the alignment time.} \textcolor{black}{Around 300 to 500 input unique sequences with an average sequence length of 7 to 10 events can be aligned relatively fast ($<$ 10s).} \textcolor{black}{To achieve real time interaction, Algs.~1 and 2 are precomputed, except for the Score and Simplify steps that are calculated on the fly as the value of $k$ and $I_\tau$ change.} \textcolor{black}{Further details about the implementation and time analysis can be found in Appendices A and B.}

\section{Analytic tasks}

The following analytic tasks were defined through a series of interviews and feedback sessions with three groups of stakeholders in the clinical domain \textcolor{black}{(emergency services, cardiac intensive care, and outpatients). Our experts shared a common goal: discover a summary of the distinct clinical pathways by clustering patients following a similar journey. They wanted to understand common and deviating scenarios to} optimise the delivery of healthcare. \textcolor{black}{Their event data sequences were partially derived from \textit{processes}. For example, an ambulance does not arrive at the scene unless a call requesting such service happens first, but a patient might be prescribed a series of drugs in indistinct order. In this context, the use of soft patterns \cite{gotz2016soft} is more appropriate - as the clustering should ensure that sequences with a similar set of event types are in the same cluster, regardless of event permutations, as opposed to enforcing a strict event ordering within cluster.} \textcolor{black}{Appendix C shows domain-specific examples for the following tasks.}

\begin{itemize}[noitemsep]
    \item [\textbf{T1.}] \textbf{Explore common and deviating pathways:} help users to explore and discover \textcolor{black}{which clusterings summarize better the most common (and deviating) pathways in the data. Clusters will group sequences that share a set of event types, regardless of their order.}
    \item [\textbf{T2.}]\textbf{Interpret the sequences that constitute a cluster:} the visualization should allow users to compare \textcolor{black}{the most common event orderings (and permutations) within and across clusters using sequence alignment.}
    \item [\textbf{T3.}] \textbf{Focus the analysis on a selected set of records:} allow queries in the dataset to focus on sequences with specific characteristics.
    \item [\textbf{T4.}] \textbf{Obtain details on demand:} provide coordinated views so that users can request finer details of interesting items in the overview. Users should be able to go from the highest level of aggregation (i.e. clusters), passing through sequences grouped by their unique sequence, to individual sequences and their raw data including event timestamps and duration.
    \item [\textbf{T5.}] \textbf{Aggregate and compare context information for selected groups of records:} the system should allow to aggregate and compare data attributes (e.g. age, gender, country) for selected clusters, unique sequences, or individual sequences.
\end{itemize}

\section{Visualization system: Sequen-C}\label{section:system}

Sequen-C was designed according to the analytic tasks outlined in the previous section. \textcolor{black}{The GUI  is implemented in Java and} it is composed by three coordinated views (Fig.~\ref{fig:coordinatedViews}): \textit{multilevel overview}, \textit{unique sequence view}, and \textit{individual sequence view}; and  the \textit{attribute analysis view}.

\subsection{The multilevel overview: cluster view}

The multilevel overview (Fig.~\ref{fig:coordinatedViews}-A) shows a variable number of sequence clusters, where each cluster is visually encoded according to the data representation matrix $\alpha$ constructed using the steps Align-Score-Simplify. Users can interact with this view through two sliders to transform the horizontal and vertical level-of-detail of the overview.

\subsubsection{Visual encoding}

Event types are represented as coloured boxes with a height proportional to the number of records and color indicates the event type. Equal event types in consecutive rows are merged to reduce visual clutter. Sequences in a cluster are ordered by similarity, and gaps (--) in the alignment are encoded as spaces between events. 

The final height of a cluster is proportional to the number of records it contains, however, sometimes clusters might contain too few records in proportion to the whole dataset ending up not visible. In such cases the height is scaled up by a constant number of pixels and the cluster is surrounded by a dotted line, allowing users to identify deviating pathways \textbf{(T1)}.

Fig.~\ref{fig:clusterEncoding} shows how each of the elements in the representation matrix  contains either one or multiple event types, where an element with multiple event types corresponds to the row wise merged sub-sequences in the Simplify step. Sub-sequences contained in a single element are represented using a box divided by colored bars, where each bar is colored by event type and ordered as per the sub-sequence. This visual encoding allows to derive the original sequences forming a cluster \textbf{(T2)}. To reduce visual clutter, when the number of events in the merged sub-sequence increases, bars can be ordered by event type to show proportion, or colored in gray to show the number of merged records.

\subsubsection{Transforming the level-of-detail} 

Fig.~\ref{fig:coordinatedViews}-E shows the two sliders used to transform the level-of-detail of the overview: the \textit{cluster slider} and the \textit{information score slider}. The cluster slider transforms the vertical level-of-detail by changing the number of clusters $k$ in the range $1 \leq k \leq N$, being $N$  the number of input unique sequences. A combobox next to the cluster slider shows the current number of clusters and contains the list of alternative optimal number of clusters (\autoref{sub:optimalOverviews}), to guide users in finding a set of pathways that best summarize the data \textbf{(T1)}. Alternatively, users can break down a selected cluster into its two child sub-clusters, and so on, until a cluster with a single sequence is reached (Fig.~\ref{fig:teaser}). The information score slider transforms the horizontal level-of-detail by changing the information score threshold $I_\tau$ in the range $0 \leq I_\tau \leq 1$. 

\subsection{Unique sequence view}
This view shows the individual sequences  in the selected clusters, grouped by unique sequence \textbf{(T4)}. The sequences are visually encoded as an ordered sequence of boxes arranged horizontally and colored by event type (Fig.~\ref{fig:coordinatedViews}-B), along with their identifier and frequency. Sequences are shown without any simplification allowing the inspection of the full sequences in the selected clusters \textbf{(T2,T4)}. These sequences can be sorted by frequency or similarity, or aligned by a selected event.

\subsection{Individual sequence view}

This view shows the individual sequences (see Fig.~\ref{fig:coordinatedViews}-C) of the selections in the unique sequence view and the overview \textbf{(T4)}, along with their temporal information and raw data attributes. Following a Gantt chart approach, each individual sequence is visualized as a horizontal sequence of events, positioned along the horizontal axis according to their timestamp. A table of attributes is displayed next to the Gantt chart, where each column represents a data attribute at either individual sequence level or individual event level \textbf{(T5)}.

\subsection{Attribute analysis view}

The distribution of a data attribute can be analysed for a selected set of records \textbf{(T3)}, or compared amongst clusters and unique sequences \textbf{(T5)}. This view shows one stacked bar chart per attribute in the dataset (see Fig.~\ref{fig:coordinatedViews}-D), where a chart contains one vertical bar per value, each bar is divided in sub-bars representing series, and series are identified by a unique color. Series can be interactively hidden to focus on only one or compare a reduced number of series. Three types of charts are provided: \textbf{1) Selected data}: compares the selected data against the rest of the records in the dataset. For a given attribute, this type of bar chart shows one series colored in red for the records contained in the selected clusters or unique sequences, and another series (in grey) for the rest of data. \textbf{2) Sequence}: it plots one series for each unique sequence shown in the unique sequence view. \textbf{3) Cluster}: it compares all clusters in the overview, and assigns one series per cluster (e.g. Fig.~\ref{fig:caseStudyCHC}-B). 

\subsection{Filters and Selections}

Records can be removed from the overview by applying filters based on data attributes, frequency, date range, event occurrence; including filters by day of the week, month, or year \textbf{(T3)}. A filter is specified by an attribute, operator, and value. For example, the filter \textit{event = A} translates to \say{show only sequences that contain event A at least once}. Users can select sections of a cluster, such as events and sub-sequences, or sequences in the unique sequence view by drawing a square with the mouse. These selections are added to the unique sequence view and individual sequence view, and are plotted in the attribute analysis 
view \textbf{(T4)}.

\section{Case studies}

In line with the analytic tasks \textcolor{black}{\textbf{(T1 to T5)}}, we present two case studies using real-world datasets: CUREd and MIMIC-III (Table \ref{table:performanceTable}).

\subsection{CUREd: Analyzing emergency service calls}

\begin{figure*}
   \centering
   \includegraphics[width=\textwidth]{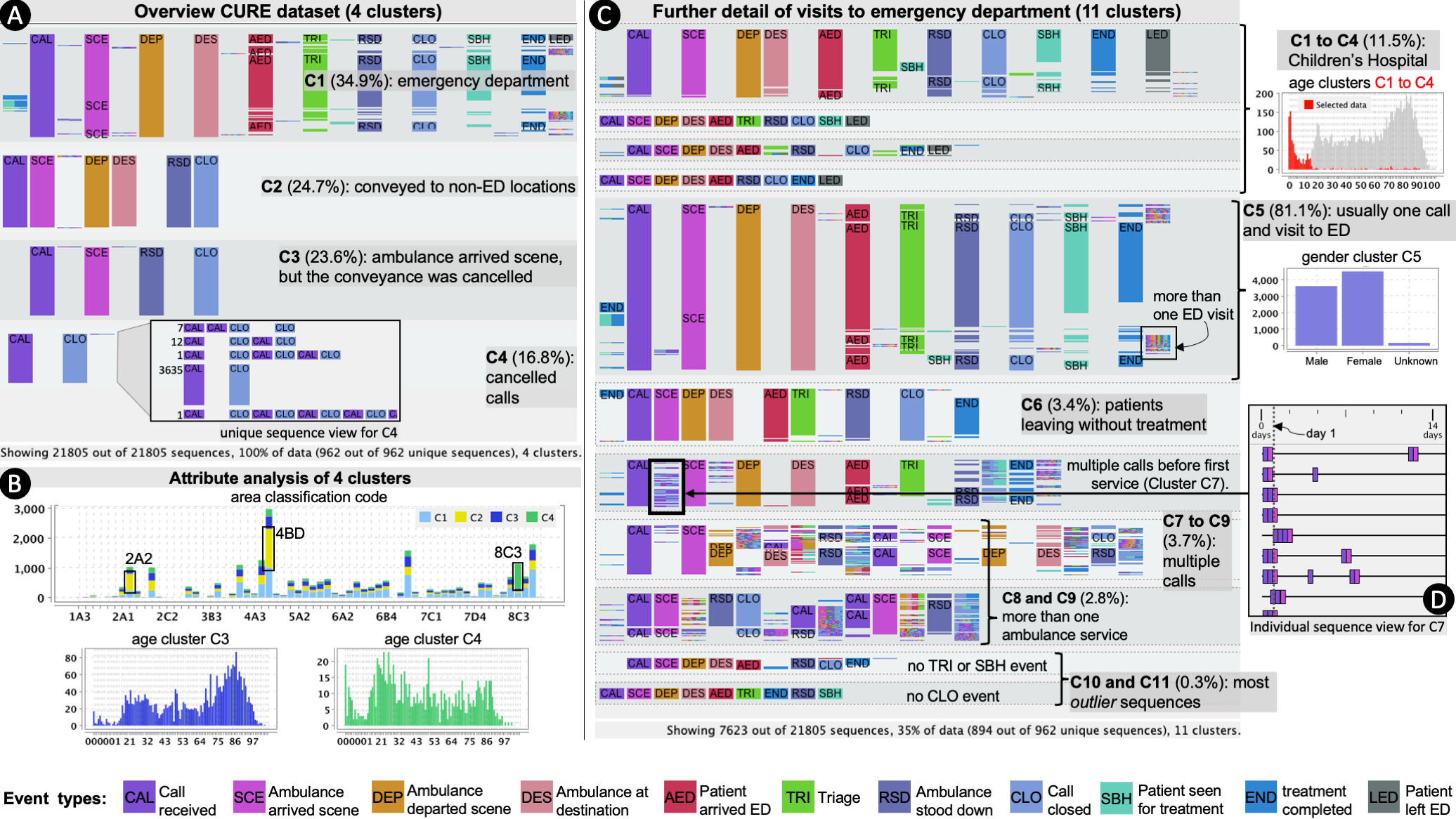}
   \caption{Multilevel overview for 21,805 patients who made calls to emergency services, obtained from the CUREd dataset.}
   \label{fig:caseStudyCHC}
\end{figure*}

The CUREd research database \cite{cureddata} contains timestamped events and demographic data related to telephone calls made to the emergency service (calls to 999 or 111), throughout Yorkshire and the Humber region. Calls can lead to different pathways, including ambulance conveyance to the Emergency department (ED) and admissions to inpatient facilities. A three month subset of the dataset was used, containing 25,243 calls relating to 21,805 unique patients, and 57 data attributes. The data were processed so that an individual sequence represents all the events of multiple calls and incidents for the same patient. We conducted an analysis session along with members of the Centre for Urgent and Emergency Care Research (CURE).

\subsubsection{Overview of main pathways}
After loading the data, the overview shows 200 clusters ordered by frequency. Clusters are well separated and present little intra-cluster variation (see Fig.~\ref{fig:coordinatedViews}). The first four clusters represent 85\% of the data and the remaining patterns are repetitions or variations of the main four. To get a coarser overview, the number of clusters is changed to 4 (see Fig.~\ref{fig:caseStudyCHC}-A). These clusters cover the pathways \textcolor{black}{\textbf{(T1)}}: Cluster C1 (34.9\%), ambulance service and attendance to the emergency department; Cluster C2 (24.7\%), ambulance service and conveyance to a hospital; Cluster C3 (23.6\%), ambulance arrives to destination but no conveyance is made. Cluster C4 (16.8\%): call closed, no ambulance service.

The attribute \textit{area classification code} is a classification based on socio-economic information derived from the postcode of the incident. Comparing the area classification code of the four clusters (see Fig.~\ref{fig:caseStudyCHC}-B), some clusters predominate more in certain area codes than others. Hospital transfers (cluster C2) happen for 43.5\% of calls coming from area code 4BD and 53.1\% of calls from area 2A2, while 72.3\% of calls from area 8C3 are in cluster C4. The age attribute (\textit{amb\_callage}) indicates that cluster C4 is more common amongst younger people, meaning that calls from area 8C3 or people in their 20s can usually be handled without an ambulance service resource attending. Conversely, cluster C3 is more common for people in their 80s (see Fig.~\ref{fig:caseStudyCHC}-B). According to the attribute \textit{symptom}: 59.7\% of calls due to chest pain end in an attendance to the emergency department (cluster C1) whereas 45.3\% of calls due to a psychiatric incident fall in cluster C4 \textcolor{black}{\textbf{(T5)}}. To further explore cluster C1, a filter is applied to show only sequences containing at least one emergency department event \textcolor{black}{\textbf{(T3)}}.

\subsubsection{Calls leading to the emergency department}
The analyst chooses 11 as the number of clusters, suggested by the system as one of the optimal number of clusters \textcolor{black}{\textbf{(T1)}}, and orders clusters by similarity. Fig.~\ref{fig:caseStudyCHC}-C shows that the first six clusters (C1 to C6) contain about 96\% of the filtered data and represent patients with usually only one call to the emergency service. These six clusters categorize visits depending on whether the visit to the emergency department is either followed by a triage event (TRI), seen by a health professional (SBH) to arrange treatment, or both \textcolor{black}{\textbf{(T2)}}. According to the attribute \textit{attendance disposal} (i.e. how the visit was concluded), in most cases where a triage event is not followed by an SBH event (clusters C3 and C6) is because the patient \textit{left the department before being treated}. Cluster C5 contains the highest percentage of data (81.1\%) with 6180 individual sequences; its attribute analysis indicates that 52\% of these patients were admitted to a hospital bed and that this cluster is slightly more common amongst women (Fig.~\ref{fig:caseStudyCHC}-C).

Interestingly, the event \say{patient left emergency department} (LED) is only present in clusters C1 to C4. The attribute analysis showed that the age of these patients go from 0 to 15 years old, and that the Children's hospital is the only ambulance destination (Fig.~\ref{fig:caseStudyCHC}-C) \textcolor{black}{\textbf{(T5)}}. The absence of the event LED in other hospitals might be due to a different configuration in the event log capturing system.

Clusters C7 to C9 suggest that about 3.7\% of patients have called the emergency service twice or more. In order to find out how many calls were made before the first ambulance service resource attended: clusters C7 to C9 are aligned by the first occurrence of the events CAL (call) and SCE (ambulance arrived scene) (see Fig.~\ref{fig:caseStudyCHC}-C). Sequences in cluster C7 contain many more calls before the first SCE event \textcolor{black}{\textbf{(T2)}}; which suggests that these patients had to \say{try} more times to get an ambulance service for the first time compared to clusters C8 and C9. To investigate whether these multiple calls were made in the same day the ambulance service was provided, cluster C7 was added to the individual sequence view \textcolor{black}{\textbf{(T4)}}. Fig.~\ref{fig:caseStudyCHC}-D shows that most of these multiple calls were made in the same day and therefore relate to the same incident. Individual scenarios in this cluster need further exploring.

\subsection{MIMIC-III}
The MIMIC-III database \cite{johnson2016mimic} contains data for 58,976 patient admissions to acute and critical care units at a tertiary hospital, organised in 26 tables containing demographic data and timestamped clinical events from admission to discharge. In this case study, an individual sequence represents all the events for a single admission, obtained from the admissions, transfers, and prescriptions tables. This case study was developed in collaboration with a consultant cardiologist (i.e. the analyst). A query was added to show patients with a primary or secondary diagnosis of Atrial Fibrillation (AF) (code 42731 in the DIAGNOSES\_ICD table). The subset data contained 1,425 patient admissions, and 448 event types, from which 438 are types of prescriptions.

\subsubsection{Overview of care unit and prescription patterns}
\begin{figure}[t]
   \centering
   \includegraphics[width=\columnwidth]{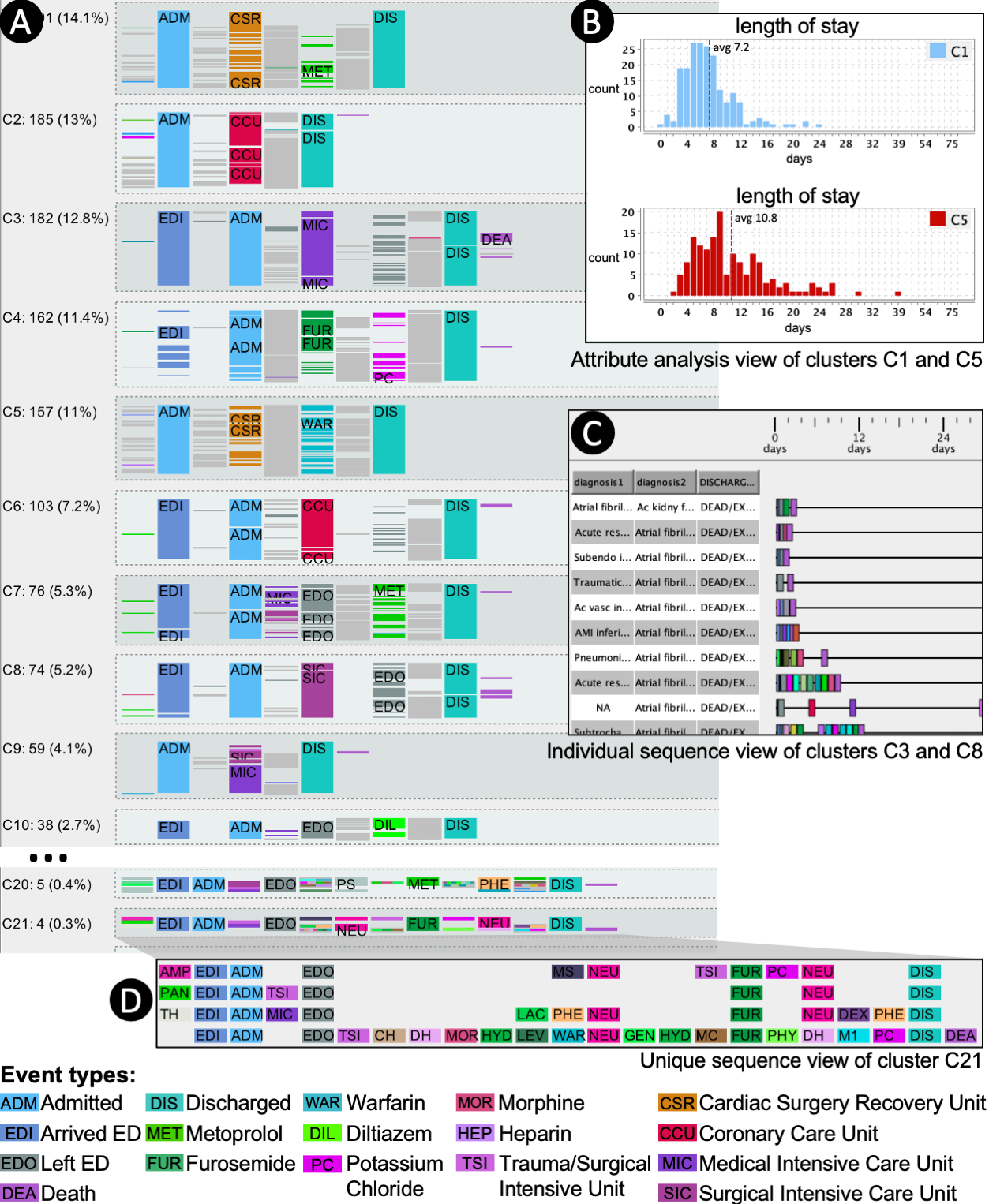}
   \caption{Overview of 1,425 admissions of patients with a first or second diagnosis of Atrial Fibrillation, obtained from the MIMIC-III dataset.}
   \label{fig:caseStudyMimic}
\end{figure}

The overview of the 1,425 individual sequences was explored by the analyst and us, selecting different number of clusters and breaking down into more granular detail within interesting clusters, the final overview shows 33 sequence clusters. The commonest care units for patients with AF were the: \textit{Cardiac Surgery Recovery Unit} (CSRU), \textit{Coronary Care Unit} (CCU), \textit{Medical Intensive Care Unit} (MICU), and \textit{Surgical Intensive Care Unit} (SICU). By default clusters were ordered by similarity. The first 18 clusters, comprising 55\% of all the admissions, started attendance to the Emergency Department (ED) with subsequent transfer to an inpatient care unit \textcolor{black}{\textbf{(T2)}}. Clusters were ordered by frequency. Fig.~\ref{fig:caseStudyMimic}-A shows that, in general, the selected clustering either groups patients sharing a specific drug but admitted to different care units (e.g. Furosemide and Potassium Chloride predominate in cluster C4), or patients admitted to a specific care unit that can be sub-divided into different treatments (e.g. admission to CSRU in cluster C1) \textcolor{black}{\textbf{(T1)}}. Clusters with less than 1\% of frequency represent more exceptional \textit{outlier} scenarios with less intra-cluster variation and very similar set of drugs (Fig.~\ref{fig:caseStudyMimic}-D) \textbf{(T2)}.

\subsubsection{Comparing attributes across clusters}

Focusing on the main clusters (C1 to C9), for each care unit, there is a cluster of patients admitted directly to that care unit and a second cluster of patients passing through ED before being transferred to that unit (e.g. clusters C2 and C6). As observed in Clusters C1 and C5, this is different in the case of the CSRU unit, where most patients do not pass through ED first \textcolor{black}{\textbf{(T2)}}. Most of the patients in cluster C1 were treated with Metoprolol; whereas most in cluster C5 were treated with Warfarin. To inspect these patients, clusters C1 and C5 were analyzed in the attribute analysis view \textcolor{black}{\textbf{(T5)}}, Fig.~\ref{fig:caseStudyMimic}-B shows that patients in cluster C5 tend to have longer lengths of stay (11 days in average) compared to cluster C1 (7 days in average). The analyst mentioned that this is likely to be associated with the requirement for careful dose \textit{titration} with Warfarin and that \say{such observations could be helpful to healthcare planning; outpatient dosing could justifiably be targeted at this cluster to reduce length of stay and free up valuable hospital beds}.

\subsubsection{Details on-demand for records of interest}

The analyst was curious to explore the clusters showing a higher mortality (C3 and C8) and their relation with a first or second diagnosis of AF. C3 and C8, containing admissions to MICU and SICU respectively, were added to the individual sequence view \textcolor{black}{\textbf{(T4)}}. Attributes \textit{diagnosis1}, \textit{diagnosis2}, and \textit{discharge\_location} were added to the Gantt chart table, and sequences were ordered by \textit{diagnosis2} and \textit{discharge\_location}. This showed that, for C3 and C8, there is a significant higher number of deaths when AF is a second diagnosis compared to when it is a first diagnosis (Fig.~\ref{fig:caseStudyMimic}-C). This might be probably because first non atrial fibrillation diagnoses might be more serious conditions.

\subsection{Domain expert overall feedback}

The domain experts (E1, E2, E3) confirmed the plausibility of the findings or considered that they required further investigation. In a separate session, we asked them to provide feedback about the usefulness of Sequen-C and the vertical and horizontal level-of-detail controls. \textbf{E1} said that the clustering suggested by the system was useful \textit{\say{to discover how patient journeys differ beyond what we would expect}}. \textbf{E3} stated that this type of analysis \textit{\say{can help to better understand clinical workflow data to improve services}}. \textbf{E2} particularly liked the functionality of exploring the attributes for a selected cluster. Experts found that the vertical level-of-detail control allows one to \textit{\say{rapidly and intuitively manipulate the granularity [of the visualization]}} (\textbf{E3}) and that increasing the number of clusters is useful when looking for outliers \textbf{(E1)}.

\textbf{E1} indicated that \textit{\say{often the high frequency events [in the cluster view] are the most important}} as these pathways would usually be the ones targeted for interventions to improve outcomes, however sometimes the interest could be in rare scenarios \textit{\say{that may have led to an adverse event}}. \textbf{E1} found useful having the flexibility of adjusting the horizontal level-of-detail to focus on both scenarios. Expert \textbf{E2} said that the horizontal slider was useful to \textit{\say{cluster noise}}. Experts mentioned that it was sometimes confusing to know which sub-sequences had \textit{collapsed} when the horizontal slider was changed and that they would prefer changing $I_\tau$ in smaller steps. Experts \textbf{E1} and \textbf{E3} mentioned that they would like to continue using the system to link additional data to expand the pathways and explore longer term clinical outcomes data.

\section{Discussion and Conclusion}
We presented a technique to create and explore a multilevel overview of event sequences through hierarchical aggregation and a novel cluster representation Align-Score-Simplify. Users can interactively transform the overview from coarse to fine \textit{vertical} or \textit{horizontal level-of-detail}, allowing the exploration of sequence clusterings (vertical) and the interactive summarization of clusters (horizontal). Moreover, we provide a set of optimal number of clusters that represent alternative good overviews to the default one. The visual analytics framework, Sequen-C, allows multilevel and details-on-demand exploration, and the inspection of data attributes at cluster, unique sequence, or individual record level.

Two case studies using real-world datasets were presented, \textcolor{black}{where findings obtained with Sequen-C were} validated with domain experts. One of the experts mentioned that traditionally, obtaining this type of findings would \textcolor{black}{involve} several meetings \textcolor{black}{between} stakeholders and analysts\textcolor{black}{, while with Sequen-C, they were able to interactively answer multiple} questions about the patterns \textcolor{black}{in a single session, consequently saving} them valuable time. The analysts in the CUREd case study mentioned that these analyses are helping them gain \say{insights into which calls are likely to need a hospital transfer and which may benefit from a different response} and that \say{knowing these patterns might help assist decision making for call handlers}.

\textcolor{black}{Although the presented case studies include} a relatively high number of event types (MIMIC-III, 448 event types) and a high number of individual sequences (CUREd, 21,805 individual sequences), \textcolor{black}{the visualization presents scalability limitations. Events are color-coded, making it difficult to distinguish more than a certain number of event types via color (e.g. 12), and navigating long lists of event types (e.g. $>20$) is complicated. In the latter case, a hierarchy of event types could facilitate this interaction.}

\textcolor{black}{An advantage of the cluster representation is that} the event types in the simplified sub-sequences are explicitly encoded, which allows to understand variability and in some cases derive the original sequences. \textcolor{black}{However, the Align-Score-Simplify approach presents three main limitations. 1)} Future work is needed to determine the optimal value of the information score $I_\tau$. \textcolor{black}{2) With increased number of event types, the representation of merged sub-sequences suffer from visual clutter.} Future work is needed to provide alternative designs \textcolor{black}{that better summarize simplified sub-sequences and improve the interpretation of information loss.} \textcolor{black}{3) The alignment step of the approach is too time consuming, it is} affected by the choice of gap and substitution costs, and it would benefit from an event type categorization (e.g. care units, prescriptions), so that events can be aligned based on their meaning rather than the name of the event type. \textcolor{black}{Alternative alignment methods could be proposed (e.g. based on the longest common sub-sequence)}.

We have created a flexible overview that allows users to uncover hidden insights. However, the current technique still depends, at some level, \textcolor{black}{on the nature of the dataset, and} the knowledge and hypotheses of domain experts. \textcolor{black}{Future work is needed to validate the current technique in other domains besides the clinical one. Lines of future research are to consider in the clustering aspects such as event type importance, event ordering (with strategies to avoid noise)} and other data attributes, and provide a visual cluster representation that encodes such attributes.

\acknowledgments{ The authors wish to thank CONACYT and The Health Foundation (PathAnalyse project) for supporting this work, \textcolor{black}{and the anonymous reviewers for their valuable comments}. CUREd Research Database is an independent project funded by the National Institute for Health Research (NIHR) Applied Research Collaboration Yorkshire and Humber (NIHR200166). The views expressed in this publication are those of the author(s) and not necessarily those of NIHR or the Department of Health and Social Care.  Paul D Morris (PD) was funded by the Wellcome Trust [214567/Z/18/Z]. For the purpose of Open Access, PD has applied a CC BY public copyright licence to any Author Accepted Manuscript version arising from this submission.}

\bibliographystyle{abbrv}

\bibliography{mybibliography}

\begin{thebibliography}{10}

\bibitem{aggarwal_reddy_2014}
C.~C. Aggarwal and C.~K. Reddy.
\newblock {\em Data clustering: algorithms and applications}.
\newblock CRC Press, 2014.

\bibitem{bose2010trace}
R.~J.~C. Bose and W.~van~der Aalst.
\newblock Trace alignment in process mining: opportunities for process
  diagnostics.
\newblock In {\em International Conference on Business Process Management},
  pages 227--242. Springer, 2010.

\bibitem{bouarfa2012workflow}
L.~Bouarfa and J.~Dankelman.
\newblock Workflow mining and outlier detection from clinical activity logs.
\newblock {\em Journal of biomedical informatics}, 45(6):1185--1190, 2012.

\bibitem{cadez2003model}
I.~Cadez, D.~Heckerman, C.~Meek, P.~Smyth, and S.~White.
\newblock Model-based clustering and visualization of navigation patterns on a
  web site.
\newblock {\em Data mining and knowledge discovery}, 7(4):399--424, 2003.

\bibitem{cappers2018exploring}
B.~C. Cappers and J.~J. van Wijk.
\newblock Exploring multivariate event sequences using rules, aggregations, and
  selections.
\newblock {\em IEEE Transactions on Visualization and Computer Graphics},
  (1):532--541, 2018.

\bibitem{chen2018sequence}
Y.~Chen, P.~Xu, and L.~Ren.
\newblock Sequence synopsis: Optimize visual summary of temporal event data.
\newblock {\em IEEE Transactions on Visualization and Computer Graphics},
  24(1):45--55, 2018.

\bibitem{di2020sequence}
S.~Di~Bartolomeo, Y.~Zhang, F.~Sheng, and C.~Dunne.
\newblock Sequence braiding: Visual overviews of temporal event sequences and
  attributes.
\newblock {\em IEEE Transactions on Visualization and Computer Graphics}, 2020.

\bibitem{du2016coping}
F.~Du, B.~Shneiderman, C.~Plaisant, S.~Malik, and A.~Perer.
\newblock Coping with volume and variety in temporal event sequences:
  Strategies for sharpening analytic focus.
\newblock {\em IEEE Transactions on Visualization and Computer Graphics},
  23(6):1636--1649, 2016.

\bibitem{elmqvist2009hierarchical}
N.~Elmqvist and J.-D. Fekete.
\newblock Hierarchical aggregation for information visualization: Overview,
  techniques, and design guidelines.
\newblock {\em IEEE Transactions on Visualization and Computer Graphics},
  16(3):439--454, 2009.

\bibitem{fails2006visual}
J.~A. Fails, A.~Karlson, L.~Shahamat, and B.~Shneiderman.
\newblock A visual interface for multivariate temporal data: Finding patterns
  of events across multiple histories.
\newblock In {\em 2006 IEEE Symposium On Visual Analytics Science And
  Technology}, pages 167--174. IEEE, 2006.

\bibitem{feng1987progressive}
D.-F. Feng and R.~F. Doolittle.
\newblock Progressive sequence alignment as a prerequisite to correct
  phylogenetic trees.
\newblock {\em Journal of molecular evolution}, 25(4):351--360, 1987.

\bibitem{goodstadt2001chroma}
L.~Goodstadt and C.~P. Ponting.
\newblock {CHROMA}: consensus-based colouring of multiple alignments for
  publication.
\newblock {\em Bioinformatics}, 17(9):845--846, 2001.

\bibitem{gotz2016soft}
D.~Gotz.
\newblock Soft patterns: Moving beyond explicit sequential patterns during
  visual analysis of longitudinal event datasets.
\newblock In {\em Proceedings of the IEEE VIS 2016 Workshop on Temporal \&
  Sequential Event Analysis}, 2016.

\bibitem{gotz2014decisionflow}
D.~Gotz and H.~Stavropoulos.
\newblock {DecisionFlow}: Visual analytics for high-dimensional temporal event
  sequence data.
\newblock {\em IEEE Transactions on Visualization and Computer Graphics},
  20(12):1783--1792, 2014.

\bibitem{gotz2011visual}
D.~Gotz, J.~Sun, N.~Cao, and S.~Ebadollahi.
\newblock Visual cluster analysis in support of clinical decision intelligence.
\newblock In {\em AMIA Annual Symposium Proceedings}, volume 2011, page 481.
  American Medical Informatics Association, 2011.

\bibitem{gotz2019visual}
D.~Gotz, J.~Zhang, W.~Wang, J.~Shrestha, and D.~Borland.
\newblock Visual analysis of high-dimensional event sequence data via dynamic
  hierarchical aggregation.
\newblock {\em IEEE Transactions on Visualization and Computer Graphics},
  26(1):440--450, 2019.

\bibitem{guo2018visual}
S.~Guo, Z.~Jin, D.~Gotz, F.~Du, H.~Zha, and N.~Cao.
\newblock Visual progression analysis of event sequence data.
\newblock {\em IEEE Transactions on Visualization and Computer Graphics},
  25(1):417--426, 2018.

\bibitem{guo2017eventthread}
S.~Guo, K.~Xu, R.~Zhao, D.~Gotz, H.~Zha, and N.~Cao.
\newblock Eventthread: Visual summarization and stage analysis of event
  sequence data.
\newblock {\em IEEE transactions on visualization and computer graphics},
  24(1):56--65, 2017.

\bibitem{johnson2016mimic}
A.~E. Johnson, T.~J. Pollard, L.~Shen, and other.
\newblock {MIMIC-III}, a freely accessible critical care database.
\newblock {\em Scientific data}, 3(1):1--9, 2016.

\bibitem{kaufman2009finding}
L.~Kaufman and P.~J. Rousseeuw.
\newblock {\em Finding groups in data: an introduction to cluster analysis},
  volume 344.
\newblock John Wiley \& Sons, 2009.

\bibitem{cureddata}
M.~Kuczawski, T.~Stone, and S.~Mason.
\newblock {CUREd}: Creating a research database to improve urgent and emergency
  care system research.
\newblock In {\em EUSEM Abstracts. Prague.}, page 512, 2019.

\bibitem{kwon2020dpvis}
B.~C. Kwon, V.~Anand, K.~A. Severson, S.~Ghosh, Z.~Sun, B.~I. Frohnert,
  M.~Lundgren, and K.~Ng.
\newblock Dpvis: Visual analytics with hidden markov models for disease
  progression pathways.
\newblock {\em IEEE transactions on visualization and computer graphics}, 2020.

\bibitem{kwon2016peekquence}
B.~C. Kwon, J.~Verma, and A.~Perer.
\newblock Peekquence: Visual analytics for event sequence data.
\newblock In {\em ACM SIGKDD 2016 Workshop on Interactive Data Exploration and
  Analytics}, volume~1, 2016.

\bibitem{lee2003generating}
C.~Lee.
\newblock Generating consensus sequences from partial order multiple sequence
  alignment graphs.
\newblock {\em Bioinformatics}, 19(8):999--1008, 2003.

\bibitem{levenshtein1966binary}
V.~I. Levenshtein.
\newblock Binary codes capable of correcting deletions, insertions, and
  reversals.
\newblock In {\em Soviet physics doklady}, volume~10, pages 707--710, 1966.

\bibitem{li2020ssrdvis}
C.~Li, X.~Dong, W.~Liu, S.~Sheng, and A.~Qian.
\newblock Ssrdvis: Interactive visualization for event sequences summarization
  and rare detection.
\newblock {\em Journal of Visualization}, 23(1):171--184, 2020.

\bibitem{liu2017coreflow}
Z.~Liu, B.~Kerr, M.~Dontcheva, et~al.
\newblock {CoreFlow}: Extracting and visualizing branching patterns from event
  sequences.
\newblock In {\em Computer Graphics Forum}, volume~36, pages 527--538, 2017.

\bibitem{liu2017patterns}
Z.~Liu, Y.~Wang, M.~Dontcheva, M.~Hoffman, S.~Walker, and A.~Wilson.
\newblock Patterns and sequences: Interactive exploration of clickstreams to
  understand common visitor paths.
\newblock {\em IEEE Transactions on Visualization and Computer Graphics},
  23(1):321--330, 2017.

\bibitem{makanju2008logview}
A.~Makanju, S.~Brooks, A.~N. Zincir-Heywood, and E.~E. Milios.
\newblock Logview: Visualizing event log clusters.
\newblock In {\em 2008 Sixth Annual Conference on Privacy, Security and Trust},
  pages 99--108. IEEE, 2008.

\bibitem{monroe2013temporal}
M.~Monroe, R.~Lan, H.~Lee, C.~Plaisant, and B.~Shneiderman.
\newblock Temporal event sequence simplification.
\newblock {\em IEEE Transactions on Visualization and Computer Graphics},
  19(12):2227--2236, 2013.

\bibitem{monroe2013challenges}
M.~Monroe, R.~Lan, J.~Morales~del Olmo, B.~Shneiderman, C.~Plaisant, and
  J.~Millstein.
\newblock The challenges of specifying intervals and absences in temporal
  queries: A graphical language approach.
\newblock In {\em Proceedings of the SIGCHI Conference on Human Factors in
  Computing Systems}, pages 2349--2358, 2013.

\bibitem{nguyen2019vasabi}
P.~H. Nguyen, R.~Henkin, S.~Chen, N.~Andrienko, G.~Andrienko, O.~Thonnard, and
  C.~Turkay.
\newblock Vasabi: Hierarchical user profiles for interactive visual user
  behaviour analytics.
\newblock {\em IEEE Transactions on Visualization and Computer Graphics},
  26(1):77--86, 2019.

\bibitem{perer2014frequence}
A.~Perer and F.~Wang.
\newblock Frequence: interactive mining and visualization of temporal frequent
  event sequences.
\newblock In {\em Proceedings of the 19th international conference on
  Intelligent User Interfaces}, pages 153--162, 2014.

\bibitem{plaisant2003lifelines}
C.~Plaisant, R.~Mushlin, A.~Snyder, J.~Li, D.~Heller, and B.~Shneiderman.
\newblock Lifelines: using visualization to enhance navigation and analysis of
  patient records.
\newblock In {\em The craft of information visualization}, pages 308--312.
  Elsevier, 2003.

\bibitem{rousseeuw1987silhouettes}
P.~J. Rousseeuw.
\newblock Silhouettes: a graphical aid to the interpretation and validation of
  cluster analysis.
\newblock {\em Journal of computational and applied mathematics}, 20:53--65,
  1987.

\bibitem{shneiderman2003eyes}
B.~Shneiderman.
\newblock The eyes have it: A task by data type taxonomy for information
  visualizations.
\newblock In {\em The Craft of Information Visualization}, pages 364--371.
  Elsevier, 2003.

\bibitem{stragier2019data}
J.~Stragier, G.~Vandewiele, P.~Coppens, F.~Ongenae, W.~Van~den Broeck,
  F.~De~Turck, and L.~De~Marez.
\newblock Data mining in the development of mobile health apps: Assessing
  in-app navigation through markov chain analysis.
\newblock {\em Journal of medical Internet research}, 21(6):e11934, 2019.

\bibitem{scikit-bio}
{The scikit-bio development team}.
\newblock scikit-bio: A bioinformatics library for data scientists, students,
  and developers, 2020.

\bibitem{ukkonen1992approximate}
E.~Ukkonen.
\newblock Approximate string-matching with q-grams and maximal matches.
\newblock {\em Theoretical computer science}, 92(1):191--211, 1992.

\bibitem{van2014stringdist}
M.~P. Van~der Loo.
\newblock The stringdist package for approximate string matching.
\newblock {\em The R Journal}, 6(1):111--122, 2014.

\bibitem{vrotsou2009activitree}
K.~Vrotsou, J.~Johansson, and M.~Cooper.
\newblock Activitree: Interactive visual exploration of sequences in
  event-based data using graph similarity.
\newblock {\em IEEE Transactions on Visualization and Computer Graphics},
  15(6):945--952, 2009.

\bibitem{wang2016unsupervised}
G.~Wang, X.~Zhang, S.~Tang, H.~Zheng, and B.~Y. Zhao.
\newblock Unsupervised clickstream clustering for user behavior analysis.
\newblock In {\em Proceedings of the 2016 CHI Conference on Human Factors in
  Computing Systems}, pages 225--236. ACM, 2016.

\bibitem{wang2003segid}
L.~Wang and Y.~Xu.
\newblock {SEGID}: Identifying interesting segments in (multiple) sequence
  alignments.
\newblock {\em Bioinformatics}, 19(2):297--298, 2003.

\bibitem{wei2012visual}
J.~Wei, Z.~Shen, N.~Sundaresan, and K.-L. Ma.
\newblock Visual cluster exploration of web clickstream data.
\newblock In {\em 2012 IEEE Conference on Visual Analytics Science and
  Technology (VAST)}, pages 3--12. IEEE, 2012.

\bibitem{wongsuphasawat2012exploring}
K.~Wongsuphasawat and D.~Gotz.
\newblock Exploring flow, factors, and outcomes of temporal event sequences
  with the {Outflow} visualization.
\newblock {\em IEEE Transactions on Visualization and Computer Graphics},
  18(12):2659--2668, 2012.

\bibitem{wongsuphasawat2011lifeflow}
K.~Wongsuphasawat, J.~A. Guerra~G{\'o}mez, C.~Plaisant, T.~D. Wang,
  M.~Taieb-Maimon, and B.~Shneiderman.
\newblock {LifeFlow}: visualizing an overview of event sequences.
\newblock In {\em Proceedings of the SIGCHI conference on human factors in
  computing systems}, pages 1747--1756. ACM, 2011.

\bibitem{wongsuphasawat2014using}
K.~Wongsuphasawat and J.~Lin.
\newblock Using visualizations to monitor changes and harvest insights from a
  global-scale logging infrastructure at twitter.
\newblock In {\em 2014 IEEE Conference on Visual Analytics Science and
  Technology (VAST)}, pages 113--122. IEEE, 2014.

\bibitem{zgraggen2015s}
E.~Zgraggen, S.~M. Drucker, D.~Fisher, and R.~DeLine.
\newblock (s|qu)eries: Visual regular expressions for querying and exploring
  event sequences.
\newblock In {\em Proceedings of the 33rd Annual ACM Conference on Human
  Factors in Computing Systems}, pages 2683--2692, 2015.

\bibitem{zhang2018idmvis}
Y.~Zhang, K.~Chanana, and C.~Dunne.
\newblock Idmvis: Temporal event sequence visualization for type 1 diabetes
  treatment decision support.
\newblock {\em IEEE transactions on visualization and computer graphics},
  25(1):512--522, 2018.

\bibitem{zhang2019evaluating}
Y.~Zhang, S.~Di~Bartolomeo, F.~Sheng, H.~Jimison, and C.~Dunne.
\newblock Evaluating alignment approaches in superimposed time-series and
  temporal event-sequence visualizations.
\newblock In {\em 2019 IEEE Visualization Conference (VIS)}, pages 1--5. IEEE,
  2019.

\bibitem{zhao2015matrixwave}
J.~Zhao, Z.~Liu, M.~Dontcheva, et~al.
\newblock {MatrixWave}: Visual comparison of event sequence data.
\newblock In {\em Proceedings of the 33rd Annual ACM Conference on Human
  Factors in Computing Systems}, pages 259--268, 2015.

\bibitem{zhou2017evaluation}
M.~Zhou, S.~Yang, X.~Li, S.~Lv, S.~Chen, I.~Marsic, R.~A. Farneth, and R.~S.
  Burd.
\newblock Evaluation of trace alignment quality and its application in medical
  process mining.
\newblock In {\em 2017 IEEE International Conference on Healthcare Informatics
  (ICHI)}, pages 258--267. IEEE, 2017.

\end{thebibliography}


\end{document}